\input{aipcheck}

\newcommand{\Msun}{\mathrm{M_{\odot}}}

\documentclass[
    ,final            
    ,dvips            
  ]
  {aipproc}

\layoutstyle{6x9}

\usepackage{amssymb}

\begin{document}

\title{Radiation-Driven Outflows in Active Galactic Nuclei}


\author{Daniel Proga}{
  address={Department of Physics and Astronomy, University of Nevada Las Vegas,
Box~454002, 4505~Maryland Pkwy, Las Vegas, NV 891541-4002}
}

\author{Ryuichi Kurosawa}{
  address={Department of Physics and Astronomy, University of Nevada Las Vegas,
  Box~454002, 4505~Maryland Pkwy, Las Vegas, NV 891541-4002}, 
  altaddress={Current Address: Department of Astronomy, Cornell University, Ithaca, NY 14853-6801} 
}

\begin{abstract}
We review the results from multi-dimensional, time-dependent
simulations of gas dynamics in AGN. We will focus on two types of
outflows powered by radiation emitted from the AGN central engine: (i)
outflows driven from the innermost part of an accretion disk and (2)
outflows driven from a large-scale inflow that is likely the main
supplier of material to the central engine. We discuss the relevance
of both types of outflows to the so-called AGN feedback
problem. However, the AGN feedback should not be considered separately
from the AGN physics. Therefore, we also discuss the issue whether the
properties of the same outflows are consistent with the gas properties
in broad- and narrow-line regions.
\end{abstract}

\maketitle


\section{Introduction}

The dynamics of the gas and dust in narrow line regions (NLRs) and 
broad line regions (BLRs) in active galactic nuclei (AGNs) is driven
by gravity but radiation driving can also be important
even for sub-Eddington sources (e.g., \cite{Krolik:1999}).
Radiation driving can be due to radiation pressure, radiation heating,
or both. The radiation force can overcome gravity for sub-Eddington sources
when the gas/dust opacity is higher than the electron scattering.
The latter is usually used to define the Eddington luminosity, $L_{\rm Edd}$.
The gas opacity can be enhanced by
the scattering of photons by UV spectral lines.
Radiation pressure on spectral lines (line force)
can be significant, provided that the gas is moderately ionized and
can interact with the UV continuum through very many UV line transitions.
For highly ionized gas, line force is negligible
because of a lower concentration of ions capable of providing UV line opacity.
In the case of highly or fully ionized gas, an outflow can still be produced
if the gas heating is efficient enough for the thermal energy
to exceed the gravitational energy.

AGNs with their broad spectral energy
distributions (SEDs), are systems where both line driving and
radiation heating, in particular X-ray heating, can operate.
In fact, a wind driven from an accretion disk by line force is the most
promising hydrodynamics (HD) scenario for outflows in AGN, especially high-luminosity quasars.
In this scenario, a wind is launched  from the disk by the
local disk radiation at radii where the disk radiation is mostly emitted
in the UV
(\cite{Shlosman:1985}; \cite{Murray:1995b}).  Such a wind  is continuous and
has a mass-loss rate and velocity that are capable of explaining
the blueshifted absorption lines observed in many AGNs, if the ionization
state is suitable 
(e.g., \cite{Murray:1995b}; \cite{Proga:2000}; \cite{Proga:2004}). 
This wind scenario has a desirable feature, i.e., for the wind motive power
it relies on radiation, which is an observable quantity.
We note that line-driving can account not only for gross properties
of some AGN outflows but also for specific spectral features
observed in some quasars  
(\cite{Arav:1994};\cite{Arav:1995}; \cite{Arav:1996}).
However, not all AGN outflows can be explained by line driving
because of too low a luminosity, too high an ionization state, or both
(e.g., \cite{Proga:2004}; \cite{Chelouche:2005};
\cite{Kraemer:2005}).
Therefore, other mechanisms such as thermal and magnetic driving are
likely also important.

Theoretical models predict that X-ray heating can have profound effects on
the gas dynamics in disks. Since X-rays tend to heat low-density gas to
a temperature $T_{\rm C}\sim 10^7$~K, with which matter in an accretion disk is
expected to either puff up and form a static corona or produce
a thermal wind, depending on whether the thermal velocity exceeds the local
escape velocity, $v_{\rm esc}$ (e.g., \cite{Begelman:1983};
\cite{Ostriker:1991}; \cite{Woods:1996}; \cite{Proga:2002}).

These studies demonstrate that
radiation liberated by an accreting
disk can drive a powerful outflow from this disk.
Therefore one can ask:
Can AGN radiation drive an outflow
from anywhere other than the disk? If yes, what are the properties
of this outflow? In particular, can it explain the properties
of BLR and NLR? Can this outflow be a part of AGN feedback?
To answer these questions,
we recently applied and extended our techniques developed
to study disk outflows to model fluid dynamics
on scales comparable to the BH's gravitational radius and also on
larger scales (\cite{Proga:2007}; \cite{Proga:2008}; \cite{Kurosawa:2008};
\cite{Kurosawa:2009a}; \cite{Kurosawa:2009b}; \cite{Kurosawa:2009c}). 
In the following, we briefly summarize the main results of 
our recent work.

\section{Simulations of outflows from inflows}

In \cite{Proga:2007}, we calculated a series of models for non-rotating flows
that are under the influence of super massive BH gravity and
radiation from an accretion disk surrounding
the BH.  Generally, we used the numerical methods developed by 
\cite{Proga:2000}.
Our numerical approach allows for the self-consistent
determination of whether the flow is gravitationally captured by the BH
or driven away by thermal expansion or radiation pressure.

To compute the structure and evolution of inflows/outflows, we solve
the equations of HD:
\begin{equation}
   \frac{D\rho}{Dt} + \rho \nabla \cdot {\bf v} = 0,
\end{equation}
\begin{equation}
   \rho \frac{D{\bf v}}{Dt} = - \nabla P - \rho \nabla \Phi+
    \rho {\bf F}^{\rm rad},
\end{equation}
\begin{equation}
   \rho \frac{D}{Dt}\left(\frac{e}{ \rho}\right) = -P \nabla \cdot {\bf v}+
 \rho \cal{L},
\end{equation}
where $\rho$ is the mass density, $P$ is the gas pressure,
${\bf v}$ is the velocity, $e$ is the internal energy density,
$\cal{L}$ is the net cooling rate,
$\Phi$ is the gravitational potential, and
${\bf F}^{\rm rad}$ is the total radiation force per unit mass.

In our previous simulations, the total radiation force has two
components: one is due to electron scattering, ${\bf F}^{\rm rad, e}$
and the other is due to lines ${\bf F}^{\rm rad, l}$. The latter can
be approximated a modified version of the method developed
by \cite{Castor:1975}.  The line force at a point defined by the
position vector $\bf r$ is

\begin{equation}
{\bf F}^{rad,l}~({\bf{r}})=~\oint_{\Omega} M(t)
\left(\hat{n} \frac{\sigma_e I({\bf r},\hat{n}) d\Omega}{c} \right)
\end{equation}
where $I$ is the frequency-integrated continuum intensity in the direction
defined by the unit vector $\hat{n}$, and $\Omega$ is the solid angle
subtended by the disk and corona at the point.
The term in brackets is the electron-scattering radiation force,
$\sigma_e$ is  the mass-scattering coefficient for free electrons,
and $M(t)$ is the force multiplier -- the numerical factor which
parametrizes how much spectral lines increase the scattering
coefficient. In the Sobolev approximation, $M(t)$ is a function
of the optical depth parameter
$ t~=~{\sigma_{e} \rho v_{\rm th}}/{\left| dv_{l}/dl \right|}$,
where $v_{\rm th}$ is the thermal velocity,
and $dv_{l}/{dl}$ is the velocity gradient along the line of sight,
$\hat{n}$.
The force multiplier approaches zero when all lines become optically
thick (i.e., ${\rm lim}_{t \rightarrow \infty} M(t)=0$)
and some finite maximum value ($M_{\rm max}$) when all lines become
optically thin (i.e., ${\rm lim}_{t \rightarrow 0}=M_{\rm max}$). 
The maximum value of the
force multiplier is a function of physical parameters of the wind and
radiation field, and can be parametrized by the photoionization
parameter, $\xi$. Several studies have shown that $M_{\rm max}$ is roughly
a few thousand for gas ionized by a weak or moderate radiation field
(e.g., \cite{Castor:1975}; 
\cite{Abbott:1982}; \cite{Stevens:1990}; \cite{Gayley:1995}).
As the radiation field becomes stronger and the gas becomes more ionized
the force multiplier decreases asymptotically to zero
(i.e., ${\rm lim}_{ \xi \rightarrow 0} M_{\rm max}=$ a few $10^3$ and
${\rm lim}_{ \xi \rightarrow \infty} M_{\rm max}=0$).
$M_{\rm max}$ allows us to set
a lower limit of the luminosity at which a system can have a ``line-driven''
wind, [i.e., $L> L_E/(1+M_{\rm max})$]. We note that
using the Castor et al.'s method \cite{Castor:1975}, one can model
large scale properties of relatively smooth line-driven winds very well,
but  cannot capture
small scale effects, in particular the so-called line driven
instability 
(e.g., \cite{Lucy:1970}; \cite{Owocki:1984}; \cite{Owocki:1988}).

We adopt an adiabatic equation of state
$P~=~(\gamma-1)e$, and consider models with the adiabatic index, $\gamma=5/3$.
We performed simulations  using the Newtonian potential $\Phi$
due to the central BH (the general relativity effects can be neglected
because we consider flow dynamics relatively far from the BH).
The simulations were performed in spherical polar coordinates
assuming axial symmetry about the rotational axis of the accretion disk.

\section{Results}

For a $10^8~\Msun$ BH with an accretion luminosity of 0.6 of $L_{\rm Edd}$,
we found that a non-rotating flow settles quickly into a steady state and has two
components
(1) an equatorial inflow and
(2) a bipolar inflow/outflow with the outflow leaving the system
along the pole.
The first component is a realization of Bondi-like accretion flow. The
second component is an example of a non-radial accretion flow becoming
an outflow once it is pushed close to the rotational axis of the disk
where thermal expansion and radiation pressure can accelerate the flow
outward. The main result of these simplified calculations is that the
existence of the above two flow components is robust yet their
properties are sensitive to the geometry, SED of the radiation field,
and the outer boundaries.  In particular, the outflow power and the
degree of collimation are higher for the model with radiation
dominated by UV/disk emission than for the model with radiation
dominated by X-ray/central engine emission.  This sensitivity is
related to the fact that thermal expansion drives a weaker and wider
outflow, compared to the radiation pressure.

In \cite{Proga:2008}, we explored effects of gas rotation.  As
expected, rotation changes the geometry of the flow because the
centrifugal force prevents gas from reaching the rotational axis (see
fig.~1 in \cite{Proga:2008}).  This, in turn, reduces the mass outflow
rate because less gas is pushed toward the polar region.  We also
found that rotation can lead to fragmentation and time variability of
the outflow. As the flow fragments, cold and dense clouds form
(figs. 4 and 12 in \cite{Proga:2008}).

\begin{figure}
  \includegraphics[clip,width=0.6\textwidth]{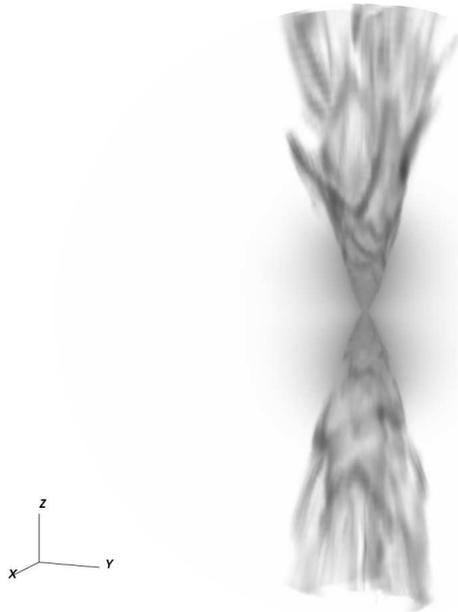}
  \caption{Three-dimensional hydrodynamical simulations of outflow
  formation via redirection of accreting gas by the strong radiation
  from an accretion disk around a super massive black hole with its
  mass $M_{\mathrm{BH}}=10^{8} \Msun$.  The infalling gas is weakly
  rotating (sub-Keplerian), and the Eddington ratio of the system is 0.6.
  The volume rendering representation of the density distributions is
  shown. The outflow morphology is bi-conical, but the flow contains
  relatively cold and dense cloud-like structures which resembles
  those observed in the NRLs of Seyfert galaxies. 
   The figure is from \cite{Kurosawa:2009a}.} 
  \label{fig:rho3d}
\end{figure}

\begin{figure}
  \includegraphics[clip,width=0.6\textwidth]{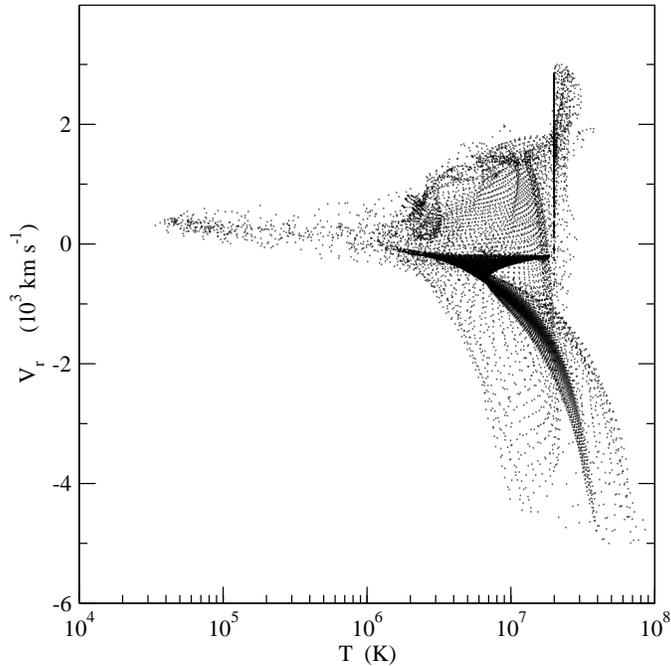}
  \caption{A scatter plot of the radial velocity ($v_{r}$)
  verses gas temperature ($T$) for the model shown in
  fig.~\ref{fig:rho3d}. A negative value of $v_{r}$ indicates an inflow.
  To avoid overcrowding, only the points on $\phi=0$ plane are shown.  A
  large fraction of gas is in outflow motion ($v_{r}>0$) for the models
  with rotation and a wide range of the temperature is associated with
  the outflowing gas. The figure is from \cite{Kurosawa:2009a}.}
  \label{fig:scatterplot}
\end{figure}

We have also started studying 3-D effects
on gas dynamics (e.g., \cite{Kurosawa:2008}; \cite{Kurosawa:2009a}).
In \cite{Kurosawa:2008}, we considered effects of radiation due to
a precessing accretion disk on a spherical cloud of gas around the disk
On the other hand, in \cite{Kurosawa:2009a}), we recalculated some models
from papers \cite{Proga:2007} and \cite{Proga:2008} in full 3-D.
Our 3-D simulations of a nonrotating gas show small yet noticeable
nonaxisymmetric small-scale features inside the outflow. However,
the outflow as a whole and the inflow do not seem to suffer from
any large-scale instability. In the rotating case, the nonaxisymmetric
features are very prominent\footnote{Sample movies of the 3-D simulations can 
be found at \url{http://www.physics.unlv.edu/~rk/research/agn_3d_rot.html}.}, 
especially in the outflow which consists of many
cold dense clouds entrained in a smoother hot flow (e.g., see 
figs.~\ref{fig:rho3d} and \ref{fig:scatterplot}). 
The 3-D outflow is nonaxisymmetric due to the shear and thermal instabilities.
Effects of gas rotations are similar in 2-D and 3-D. In particular, 
gas rotation increases the outflow thermal
energy flux, but reduces the outflow mass and kinetic energy fluxes.
In addition,
rotation leads to time variability and fragmentation of the outflow
in the radial and latitudinal directions. The collimation of the outflow
is reduced in the models with gas rotation. The main different
effect of rotation in 3-D compared to 2-D is that
the time variability in the mass
and energy fluxes is reduced in the 3-D case because of
the outflow fragmentation in the azimuthal direction.

\begin{figure}
  \includegraphics[clip,width=0.6\textwidth]{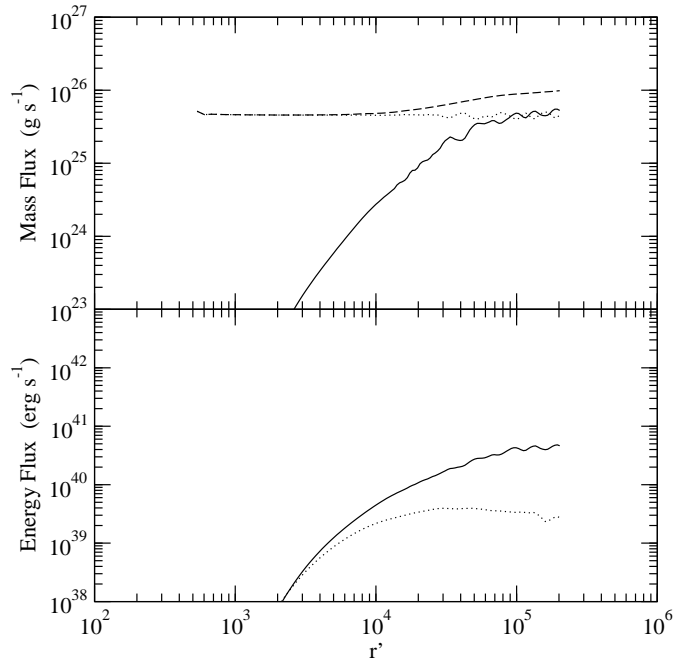}
  \caption{Mass and energy fluxes as a function of radius for the
   model shown in fig.~\ref{fig:rho3d}. The panel is subdivided into
   two parts: \emph{top} (mass flux) and \emph{bottom} (energy
   flux). In the mass flux plot, the inflow (\emph{dashed line};
   $\dot{M}_{\mathrm{in}}$), outflow (\emph{solid line};
   $\dot{M}_{\mathrm{o}}$) and net (\emph{dotted lines};
   $\dot{M}_{\mathrm{net}}$) mass fluxes, are separately plotted, as a
   function of radius. The absolute values of $\dot{M}_{\mathrm{in}}$
   and $\dot{M}_{\mathrm{net}}$ are plotted here since they are
   negative at all radii. The length scale is in units of the inner
   disk radius ($r'=r/r_{*}$ where $r_{*}=8.8\times10^{13}$~cm). In the
   energy flux plots, the kinetic energy (\emph{solid line}) and the
   thermal energy (\emph{dotted line}) fluxes are shown. The figure is
   from \cite{Kurosawa:2009a}.} 
  \label{fig:mass-energy-flux}
\end{figure}

Our recent simulations, demonstrate that AGNs can have a substantial outflow
originating from the infalling gas. Such an outflow can reduce the rate at
which matter is supplied to the central region of AGN because its mass loss rate
can be significantly higher than the mass inflow rate at small radii
(see fig.~\ref{fig:mass-energy-flux}).
For example, as little as 10\% of the inflow at large radii  can reach small
radii because 90\% of the inflow is turned into an outflow.
In general, the kinetic power dominates the thermal power at all
radii in both models with and without gas rotation; however, the
thermal power contribution is non-negligible in the models with gas
rotation (fig.~\ref{fig:mass-energy-flux}). The X-ray heating becomes
more effective in the rotating gas environment because the flows
become more non-radial and will be subjected to the direct
exposure to the strong radiation from central continuum radiation
source.  

\begin{figure}
  \includegraphics[clip, width=1.0\textwidth]{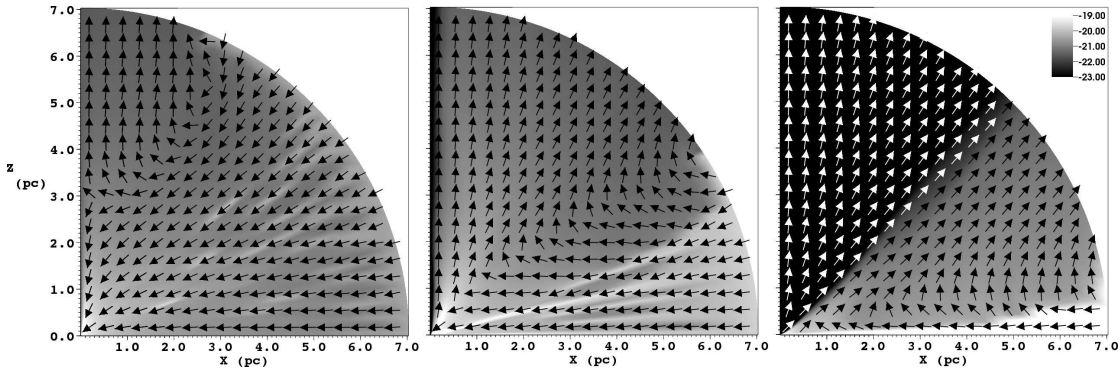}
  \caption{Examples of the density and velocity maps from a set of
  two-dimensional simulations (without the  temperature constrained at
  the outer boundary) presented in \cite{Kurosawa:2009b}. The
  density (in logarithmic scale) is over-plotted with the directions
  of poloidal velocity as arrows. The figures are placed in order of 
  increasing density at the outer boundaries ($\rho_{\mathrm{o}}$) from 
  the left to right. The self-consistently determined accretion
  luminosities (in the unit of Eddington luminosity) are $\Gamma =
  0.32, 0.71$ and $4.3$, from the  left to right panels.  
  For the low density ($\rho_{\mathrm{o}}$) and low accretion
  luminosity  model (\emph{left panel}), the outflow is  
  very narrow ($0^{\circ} \leq \theta  \lesssim 30^{\circ}$), and the
  inflow is very wide ($30^{\circ} \lesssim \theta \lesssim
  90^{\circ}$).  As the density ($\rho_{\mathrm{o}}$) and 
  accretion luminosity increase
  (\emph{middle panel}), the outflow becomes wider
  ($0^{\circ} \leq  \theta \lesssim 50^{\circ}$), whereas the inflow 
  becomes narrower ($50^{\circ} \lesssim \theta \lesssim 90^{\circ}$). For
  the very high density and accretion model (\emph{right panel}), the
  outflow occurs over a  very wide range of the polar angle ($0^{\circ} \leq
  \theta \lesssim 85^{\circ}$), and the accretion region is now confined to
  a thin equatorial wedge (the disk-wind-like solution).  The figures
  are from \cite{Kurosawa:2009c}. }
  \label{fig:rho-maps-2d}
\end{figure}

In \cite{Kurosawa:2009b}, the radiation-driven AGN outflow model
of \cite{Proga:2007} is extended by relaxing the assumption of a
constant accretion luminosity.  This allows us to determine the
accretion luminosity consistently with the mass accretion rate at the
inner boundary, and consequently the two quantities are coupled
through the radiation field.  This is an improvement toward a
more comprehensive self-consistent hydrodynamical models with
radiative feedback.
For the models with high temperature gas at large radii ($\sim10$~pc)
and high luminosities, we find a strong correlation between the
mass-outflow rate ($\dot{M}_{\mathrm{out}}$) and the luminosity ($L$).
The power law index ($q$) describing the $\dot{M}_{\mathrm{out}}$--$L$
relation is $q=2.0\left(\pm0.1\right)$, which is very similar to that
for radiation-driven stellar and disk wind models. More surprisingly,
for high density at large radii, we find steady state solutions with
the accretion luminosity {\it exceeding} the Eddington limit.  The
super-Eddington accretion proceeds in the equatorial region and is
possible because the radiation flux from the disk is significantly
reduced in the equatorial direction due to the geometrical
foreshortening effect\footnote{Sample movies of the 2-D simulations can be found
at \url{http://www.physics.unlv.edu/~rk/research/agn2d_survey.html}}
(see fig.~\ref{fig:rho-maps-2d}).

\begin{figure}
   \includegraphics[clip,width=0.6\textwidth]{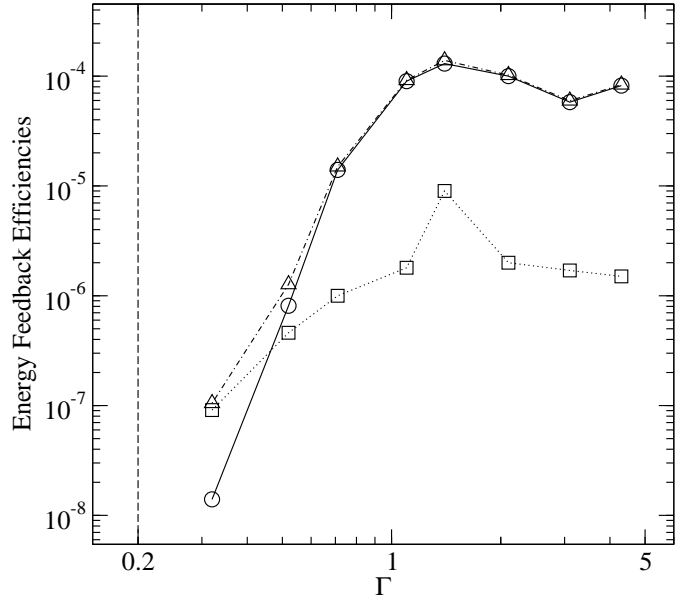}


  \caption{The efficiencies of converting the black hole accretion
  luminosity $L_{\mathrm{a}}$ to the rate of energy deposition to the
  surrounding gas plotted as a function of the Eddington ratio
  ($\Gamma$). The values are based on the 2-D simulations similar to
  those shown in fig.~\ref{fig:rho-maps-2d}.  The kinetic energy
  feedback efficiency $\epsilon_{\mathrm{k}}$ (\emph{circles}), the
  thermal energy feedback efficiency $\epsilon_{\mathrm{th}}$
  (\emph{squares}) and the total energy feedback efficiency
  $\epsilon_{\mathrm{t}}=\epsilon_{\mathrm{k}}+\epsilon_{\mathrm{th}}$
  (\emph{triangles}) are shown separately.  The maximum total energy
  feedback efficiency is $\sim10^{-4}$.  For the models with
  relatively low Eddington ratio ($\Gamma \lesssim0.4$), the thermal
  feedback is more efficient than the kinetic feedback
  ($\epsilon_{\mathrm{th}}>\epsilon_{\mathrm{k}}$).  For the models
  with relatively high Eddington ratio ($\Gamma \gtrsim 0.6$), the
  kinetic feedback is more efficient than the thermal feedback by a
  factor of $\sim10$ to $\sim100$. The model with $\Gamma=0.2$ does
  not form an outflow, and the vertical line (\emph{dashed}) at
  $\Gamma=0.2$ indicates an approximate $\Gamma$ value below which no
  outflow forms.  The flattening of the efficiencies beyond
  $\Gamma\approx1$ is caused by the transition of the inflow-outflow
  morphology to a {}``disk wind like'' configuration for the larger
  $\Gamma$ models (c.f., fig.~\ref{fig:rho-maps-2d}).  The figure
  is from \cite{Kurosawa:2009c}.
}  
  \label{fig:efficiency}

\end{figure}

Based on a set of axisymmetric simulations presented in \cite{Kurosawa:2009b}, 
we analyzed the energy, momentum, and mass feedback efficiencies due
to radiation from AGN (see \cite{Kurosawa:2009c}). 
We find that even for the strongest outflow,               
the ratio between the outflow kinetic power and the radiation power is 
very low ($\sim 10^{-4}$ at the peak; see fig.~\ref{fig:efficiency}).
One of the reasons for this relatively low efficiency of the large
scale outflows is that for large densities at large radii,
accretion proceeds in the equatorial region, and the radiation flux
from the disc is significantly reduced the geometrical foreshortening
effect, as also mentioned above.  The
coupling between the radiation and matter becomes less efficient once
the inflow-outflow morphology becomes the ``disk wind like'' (the right
panel in fig.~\ref{fig:rho-maps-2d}) because of the ``mismatch''
between the direction in which the disk radiation peaks (in the polar
direction) and the direction of the matter inflow (in the equatorial
direction).

Compared to the energy (thermal only) feedback efficiencies 
$\sim0.05$ required in the recent
cosmological and galaxy mergers simulations (e.g., \cite{Springel:2005};
\cite{Robertson:2006}; \cite{Sijacki:2007}; \cite{DiMatteo:2005}; \cite{Johansson:2009}),
our thermal energy feedback efficiency 
at the peak value is about $5\times10^{3}$ times smaller.
Our total and kinetic energy efficiencies are about $5\times10^{2}$
times smaller than the value required in cosmological simulations.
These large discrepancies might indicate a few things.
For example, our models are missing important elements. 
In particular, we do not include effects of dust which could make 
the outflows much stronger. 

On the other hand, it is also possible that the AGN feedback may not
be as effective as one might have had expected. Instead, other forms
of feedback may be more significant than the AGN feedback via radiation 
on scales between $10^{-2}$ and a few parsecs. For example, the feedback
via supernovae,  star formation processes, the strong stellar wind
from massive stars, and strong accretion disk winds (discussed
in section 1) or jets from AGN
may play more important roles. 
The last two forms of the feedback require a proper treatmentment of
magnetic field, as they may carry a significant fraction of the total
outward flux in energy and momentum.  

Finally, the AGN feedback efficiency may be indeed low, and the AGNs
take a long time to influence their environment. We note that in our
models the AGNs do not shut off the mass supply completely even at
very high luminosities ($\Gamma \gtrsim 1$).  This indicates that the AGNs can operate on a
very long time scale over which their affects can accumulate, and
eventually become significant.

\section{Conclusions}

The main conclusions from our recent simulations of radiation driven
large scale outflows in AGN include:
\begin{itemize}
  \item A significant fraction of the inflowing matter can be expelled by
  radiation heating and pressure.
  \item The non-rotating flow settles into a steady inflow/outflow solution.
  Rotation of the flow and large optical depth lead to time variability
  in the inflow/outflow solution.
  \item In the time variable flows, dense clouds form. This could be
  applicable to NLR in AGN.
  \item The large outflows are multi-temperature/phase and could explain
  warm absorbers in AGN.
  \item The mass supply rate does not appear to be limited by 
  the AGN radiation.
  \item The large-scale outflows are efficient in removing matter 
  but inefficient in carrying out energy.
  \item Radiation drives much more powerful outflows from small scales
  than from large scales.
\end{itemize}


\begin{theacknowledgments}
This work was supported by NASA through grant HST-AR-11276 from the
Space Telescope Science Institute, which is operated by the Association
of Universities for Research in Astronomy, Inc., under NASA contract
NAS5-26555.
\end{theacknowledgments}



\bibliographystyle{aipproc}   

\bibliography{dproga}

\hyphenation{Post-Script Sprin-ger}
\begin{thebibliography}{32}
\expandafter\ifx\csname natexlab\endcsname\relax\def\natexlab#1{#1}\fi
\providecommand{\enquote}[1]{``#1''}
\expandafter\ifx\csname url\endcsname\relax
  \def\url#1{\texttt{#1}}\fi
\expandafter\ifx\csname urlprefix\endcsname\relax\def\urlprefix{URL }\fi
\providecommand{\eprint}[2][]{\url{#2}}

\bibitem[{Krolik}(1999)]{Krolik:1999}
J.~H. {Krolik}, \emph{{Active galactic nuclei : from the central black hole to
  the galactic environment}}, Princeton, N.~J.~: Princeton University Press,
  1999.

\bibitem[Shlosman et~al.(1985)]{Shlosman:1985}
I.~Shlosman, P.~A. Vitello, and G.~Shaviv, \emph{\apj} \textbf{294}, 96--105
  (1985).

\bibitem[Murray et~al.(1995)]{Murray:1995b}
N.~Murray, J.~{Chiang}, S.~A. {Grossman}, and G.~M. {Voit}, \emph{\apj}
  \textbf{451}, 498 (1995).

\bibitem[{Proga} et~al.(2000)]{Proga:2000}
D.~{Proga}, J.~M. {Stone}, and T.~R. {Kallman}, \emph{\apj} \textbf{543},
  686--696 (2000).

\bibitem[Proga and Kallman(2004)]{Proga:2004}
D.~Proga, and T.~R. Kallman, \emph{\apj} \textbf{616}, 688--695 (2004).

\bibitem[{Arav} and {Begelman}(1994)]{Arav:1994}
N.~{Arav}, and M.~C. {Begelman}, \emph{\apj} \textbf{434}, 479--483 (1994).

\bibitem[{Arav} et~al.(1995)]{Arav:1995}
N.~{Arav}, K.~T. {Korista}, T.~A. {Barlow}, and {Begelman}, \emph{\nat}
  \textbf{376}, 576--578 (1995).

\bibitem[{Arav}(1996)]{Arav:1996}
N.~{Arav}, \emph{\apj} \textbf{465}, 617--630 (1996).

\bibitem[{Chelouche} and {Netzer}(2005)]{Chelouche:2005}
D.~{Chelouche}, and H.~{Netzer}, \emph{\apj} \textbf{625}, 95--107 (2005).

\bibitem[{Kraemer} et~al.(2005)]{Kraemer:2005}
S.~B. {Kraemer}, I.~M. {George}, D.~M. {Crenshaw}, J.~R. {Gabel}, T.~J.
  {Turner}, T.~R. {Gull}, J.~B. {Hutchings}, G.~A. {Kriss}, R.~F. {Mushotzky},
  H.~{Netzer}, B.~M. {Peterson}, and E.~{Behar}, \emph{\apj} \textbf{633},
  693--705 (2005).

\bibitem[{Begelman} et~al.(1983)]{Begelman:1983}
M.~C. {Begelman}, C.~F. {McKee}, and G.~A. {Shields}, \emph{\apj} \textbf{271},
  70--88 (1983).

\bibitem[{Ostriker} et~al.(1991)]{Ostriker:1991}
E.~C. {Ostriker}, C.~F. {McKee}, and R.~I. {Klein}, \emph{\apj} \textbf{377},
  593--611 (1991).

\bibitem[{Woods} et~al.(1996)]{Woods:1996}
D.~T. {Woods}, R.~I. {Klein}, J.~I. {Castor}, C.~F. {McKee}, and J.~B. {Bell},
  \emph{\apj} \textbf{461}, 767 (1996).

\bibitem[Proga et~al.(2002)]{Proga:2002}
D.~Proga, T.~R. {Kallman}, J.~E. {Drew}, and L.~E. {Hartley}, \emph{\apj}
  \textbf{572}, 382--391 (2002).

\bibitem[{Proga}(2007)]{Proga:2007}
D.~{Proga}, \emph{\apj} \textbf{661}, 693--702 (2007).

\bibitem[{Proga} et~al.(2008)]{Proga:2008}
D.~{Proga}, J.~P. {Ostriker}, and R.~{Kurosawa}, \emph{\apj} \textbf{676},
  101--112 (2008).

\bibitem[Kurosawa and Proga(2008)]{Kurosawa:2008}
R.~Kurosawa, and D.~Proga, \emph{\apj} \textbf{674}, 97--110 (2008).

\bibitem[{Kurosawa} and {Proga}(2009{\natexlab{a}})]{Kurosawa:2009a}
R.~{Kurosawa}, and D.~{Proga}, \emph{\apj} \textbf{693}, 1929--1945
  (2009{\natexlab{a}}).

\bibitem[{Kurosawa} and {Proga}(2009{\natexlab{b}})]{Kurosawa:2009b}
R.~{Kurosawa}, and D.~{Proga}, \emph{\mnras} \textbf{397}, 1791--1803
  (2009{\natexlab{b}}).

\bibitem[{Kurosawa} et~al.(2009)]{Kurosawa:2009c}
R.~{Kurosawa}, D.~{Proga}, and K.~{Nagamine}, \emph{ArXiv e-prints}  (2009),
  \eprint{0906.3739}.

\bibitem[Castor et~al.(1975)]{Castor:1975}
J.~I. Castor, D.~C. Abbott, and R.~I. Klein, \emph{\apj} \textbf{195}, 157--174
  (1975).

\bibitem[{Abbott}(1982)]{Abbott:1982}
D.~C. {Abbott}, \emph{\apj} \textbf{259}, 282--301 (1982).

\bibitem[Stevens and Kallman(1990)]{Stevens:1990}
I.~R. Stevens, and T.~R. Kallman, \emph{\apj} \textbf{365}, 321--331 (1990).

\bibitem[{Gayley}(1995)]{Gayley:1995}
K.~G. {Gayley}, \emph{\apj} \textbf{454}, 410--419 (1995).

\bibitem[{Lucy} and {Solomon}(1970)]{Lucy:1970}
L.~B. {Lucy}, and P.~M. {Solomon}, \emph{\apj} \textbf{159}, 879--893 (1970).

\bibitem[{Owocki} and {Rybicki}(1984)]{Owocki:1984}
S.~P. {Owocki}, and G.~B. {Rybicki}, \emph{\apj} \textbf{284}, 337--350 (1984).

\bibitem[Owocki et~al.(1988)]{Owocki:1988}
S.~P. Owocki, J.~I. {Castor}, and G.~B. {Rybicki}, \emph{\apj} \textbf{335},
  914--930 (1988).

\bibitem[Springel et~al.(2005)]{Springel:2005}
V.~Springel, T.~{Di Matteo}, and L.~{Hernquist}, \emph{\mnras} \textbf{361},
  776--794 (2005).

\bibitem[Robertson et~al.(2006)]{Robertson:2006}
B.~Robertson, L.~{Hernquist}, T.~J. {Cox}, T.~{Di Matteo}, P.~F. {Hopkins},
  P.~{Martini}, and V.~{Springel}, \emph{\apj} \textbf{641}, 90--102 (2006).

\bibitem[Sijacki et~al.(2007)]{Sijacki:2007}
D.~Sijacki, V.~{Springel}, T.~{di Matteo}, and L.~{Hernquist}, \emph{\mnras}
  \textbf{380}, 877--900 (2007).

\bibitem[Di~Matteo et~al.(2005)]{DiMatteo:2005}
T.~Di~Matteo, V.~Springel, and L.~Hernquist, \emph{\nat} \textbf{433}, 604--607
  (2005).

\bibitem[Johansson et~al.(2009)]{Johansson:2009}
P.~H. Johansson, T.~{Naab}, and A.~{Burkert}, \emph{\apj} \textbf{690},
  802--821 (2009).

\end{thebibliography}

\IfFileExists{\jobname.bbl}{}
 {\typeout{}
  \typeout{******************************************}
  \typeout{** Please run "bibtex \jobname" to optain}
  \typeout{** the bibliography and then re-run LaTeX}
  \typeout{** twice to fix the references!}
  \typeout{******************************************}
  \typeout{}
 }

\end{document}